\documentstyle[psfig,aps,prl]{revtex}
\begin{document}
\title{Essentially All Gaussian Two-Party Quantum States are 
{\it a priori} Nonclassical
but Classically Correlated}
\author{Paul B. Slater}
\address{ISBER, University of California, Santa Barbara, CA 93106-2150\\
e-mail: slater@itp.ucsb.edu,
FAX: (805) 893-7995}

\date{\today}

\draft
\maketitle
\vskip -0.1 cm

\begin{abstract}
Duan, Giedke, Cirac and Zoller 
and, independently, Simon have recently found  necessary and sufficient
conditions for
 the separability (classical correlation)
  of the Gaussian two-party (continuous variable) states.
Duan {\it et al} remark that their criterion is based on a ``much stronger
bound'' on the total variance of a pair of Einstein-Podolsky-Rosen-type
operators than is required simply
 by the uncertainty relation. Here, we
seek to 
formalize and test this particular assertion in both classical
and quantum-theoretic frameworks. We first attach to these
states the classical {\it a priori} probability (Jeffreys' prior),
proportional to the volume element of the Fisher information metric on the
Riemannian manifold of 
Gaussian (quadrivariate normal) probability distributions.
Then, numerical evidence (indicates that more than ninety-nine percent of 
the Gaussian two-party 
states do, in fact, meet the 
more stringent criterion for separability. We collaterally note that
the prior probability assigned to the 
classical  states, that is those having  positive 
Glauber-Sudarshan $P$-representations, is less than one-thousandth
of one percent.
We, then, seek to attach as a measure to the Gaussian two-party states, 
the volume element of the associated 
(quantum-theoretic) Bures ({\it minimal} monotone) metric.
Our several extensive analyses, then, 
 persistently yield probabilities of separability
and classicality that are, to very high orders of
accuracy, unity and zero, respectively, so the two quite distinct 
(classical and quantum-theoretic) forms of analysis
are rather remarkably consistent in their findings.
\end{abstract}

\pacs{PACS Numbers 03.67.-a, 42.50.Dv, 89.70.+c, 02.50.-r}
\section{INTRODUCTION}

\.Zyczkowski, Horodecki, Sanpera and Lewenstein (ZHSL)
\cite{zycz1} were the first to raise
and address the issue of determining the proportion of quantum states which
are --- at least, in some 
appropriate {\it a priori} or natural sense --- separable, that is,
{\it classically correlated}.
Such composite states can be approximated by convex combinations of product
states. Inseparable states are also
 termed {\it Einstein-Podolsky-Rosen correlated} 
\cite{werner}.

Though acknowledging the importance and intrinsic interest of the line of
investigation opened by ZHSL,
Slater \cite{slater1} was critical of certain aspects
of the  methodology they employed.
 He conducted alternative analyses, using normalized volume elements of
monotone metrics \cite{petzsudar,les} on the quantum states, as candidates for
prior probability distributions. It was found (although the 
various numerical calculations
were marked by considerable instabilities 
\cite[Tables 1-4]{slater1}) that the (much discussed
\cite{braunstein,hubner}) Bures (that is, {\it minimal}
 monotone) metric yielded
the highest probabilities of separability for the $2 \times 2$ and $2 \times 3$
systems. (These probabilities --- approximately, ten percent
in the $2 \times 2$ case --- were, nevertheless,
 substantially less than
those arrived at by ZHSL, counterintuitively, to their ``surprise''.
Making use of ``explicit'' formulas of Dittmann \cite{dittgood} for the
Bures metric, Slater \cite{slatgood} was then able to obtain {\it exact}
Bures probabilities of separability for certain two-{\it qubit} scenarios.)

Both the approaches of ZHSL and Slater relied upon a 
positive partial transposition
condition, shown to be sufficient for separability by Peres \cite{peres} and 
to be  necessary as well by Horodecki
\cite{horodecki} in  the specific instances of
the $2 \times 2$ and
$2 \times 3$ systems, but not necessary in higher-dimensional systems.

Duan, Griedke, Cirac and Zoller \cite{duan} (cf. \cite{lee})
 have recently found an
apparently quite distinct 
separability criterion for two-party continuous variable 
(infinite-dimensional) systems.
Such systems have served as the basis of many recent protocols for quantum
communication and computation \cite{braunnature,polkinghorne}.
Duan {\it et al}
 asserted that it was quite difficult to ascertain if the partial
transposition criterion is, in fact, 
 met. However,  Simon (in a slightly later prepublication \cite{simon},
 based on an
earlier invited talk),
was able to accomplish 
precisely this task, using a geometric interpretation
of the partial transposition operator as mirror reflection in phase
space.

The criterion of Duan {\it et al} --- which is the one we study here --- is 
based on the total variance of a pair of Einstein-Podolsky-Rosen type
operators ($ \hat{u}, \hat{v}$) of the type
\begin{equation} \hat{u} = {1 \over \sqrt{2}} (a \hat{x}_{1} \pm
{1 \over a} \hat{x}_{2}),\qquad
\hat{v} = {1 \over \sqrt{2}} (a \hat{p}_{1} \mp {1 \over
 a} \hat{p}_{2}),
\end{equation}
where $a$ is assumed to be an arbitrary positive number, and 
$\hat{x}_{j}$, 
$\hat{p}_{j}$ $(j=1,2)$ can be any local operators satisfying the commutators
$[\hat{x}_{j},\hat{p}_{j'}] = i 2 \delta_{j j'}$.
The new 
variance criterion provides a sufficient condition for entanglement of any
two-party continuous variables states. Furthermore, for those of these states
which are Gaussian,
this criterion turns out to be necessary as well.

More 
specifically, 
Duan {\it et al} transform the correlation matrix for the Gaussian 
two-party state,
using a twofold application of 
local linear unitary Bogoliubov operations (LLUBOs). They then derive
an inequality (\cite[eq. (16)]{duan}) which must be satisfied for 
separability to hold. 
(A necessary and sufficient condition for this is that the 
{\it tranformed} matrix minus the identity matrix be positive semidefinite.) 
In our numerical procedures below, we test whether
or not this inequality is met. Then, weighting the 
(systematically generated) states by 
certain classical (``Jeffreys' prior'') and quantum (Bures) measures, we obtain
estimates of the proportion of Gaussian two-party states that are separable.
As a collateral exercise, we test the 
initial untransformed matrices to 
see if they have positive $P$-representations. Again, using the classical
and quantum measures as weights, 
we estimate the proportion of Gaussian two-party 
states that are classical in character.

\section{BURES METRICS FOR CONTINUOUS VARIABLE SYSTEMS}

So, it would appear natural, in light of the previous analyses of ZHSL
\cite{zycz1}
and Slater \cite{slater1},
 to address the question of
 what proportion of the Gaussian two-party continuous variable
states are (in)separable.
Such states can always be parameterized as thermal squeezed
states \cite{mann,scutarunew}.
 Twamley \cite{twamley} has computed the Bures distance
between two (one-mode)
 thermal squeezed states and deduced the statistical distance
(Bures) metric of Braunstein and Caves \cite{braunstein}.
The computation of Twamley was largely concentrated on determining the
fidelity
\begin{equation} \label{twamfid}
F(\rho_{1},\rho_{2}) = \mbox{Tr} \sqrt{ \rho_{1}^{1/2} \rho_{2} \rho_{1}^{1/2}}
\end{equation}
(the Bures distance itself being given by $2 (1- F(\rho_{1},\rho_{2})$)
between two squeezed thermal states (having density
matrices $\rho_{1}$ and $\rho_{2}$),
where
\begin{equation}
\rho(\beta,r,\theta) = Z S(r,\theta) T(\beta) S^{\dagger}(r,\theta) \qquad
(0 \leq \beta; 0 \leq r; -\pi < \theta \leq \pi)
\end{equation}
\begin{equation}
S(r,\theta) = \exp{ (  \zeta K_{+} - \zeta^{*} K_{-})},\quad
T(\beta)=\exp{(-\beta K_{0})},\quad
\zeta= r \mbox{e}^{\mbox{i} \theta}
\end{equation}
and
\begin{equation}
K_{+} = {1 \over 2} \hat{a}^{\dagger 2}, \quad
 K_{-} = {1 \over 2} \hat{a}^2,\quad
 K_{0} = {1 \over 2} (\hat{a}^{\dagger} \hat{a} + {1 \over 2}),\quad
[K_{0},K_{\pm}] = \pm K_{\pm},\quad  [K_{-},K_{+}] =2 K_{0}.
\end{equation}
Here $S(r,\theta)$ is the one-photon squeeze operator, $\hat{a}$
 is the single-mode
annihilation operator, $Z$ is chosen so that $\mbox{Tr} (\rho) =1$, and 
$(K_{0},K_{\pm})$ are the generators of the $SU(1,1)$ group.
Twamley was able to express the fidelity 
(\ref{twamfid}) in the form \cite[eq. (24)]{twamley}
\begin{equation}
F(\rho_{1},\rho_{2}) =
 {\sqrt{2 \sinh{{\beta_{1}} \over 4} \sinh{{\beta_{2} \over 4}}} \over 
\sqrt{\sqrt{Y}} -1},
\end{equation}
where the $\beta$'s are proportional to inverse temperatures, and $Y$
corresponds to a certain (two-line) trigonometic expression.

Using the results of Twamley, Slater
 \cite[eqs. (8), (9)]{slater2} expressed the volume
element (the ``quantum Jeffreys' prior'') 
of the Bures metric in the form $f(r) g(\beta)$,
where
\begin{equation} \label{impropbures}
f(r) = \sinh{2 r}, \quad g(\beta) ={\cosh{{\beta \over 4}}
 \coth{{\beta  \over 4}}
\mbox{sech}{{\beta  \over 2}} \over 8}.
\end{equation}
(Subsequently, quantum Jeffreys' priors have been determined, as well, 
 for the 
displaced thermal states \cite{parscu} and for the displaced squeezed thermal
states \cite{kwek}. Following work of Lavenda \cite{lavenda}, 
{\it high}-temperature
expansions of these priors were considered in \cite{slatlav}.)

For the one-mode
case, Paraoanu and Scutaru have expressed the fidelity in the form
\cite[eq. (31)]{paraoanu}
\begin{equation}
 F(\rho_{1},\rho_{2}) =
 {2 \over \sqrt{ \mbox{det} (A_{1} +A_{2}) +P} -\sqrt{P}},
\end{equation}
where the $A$'s are $2 \times 2$ correlation matrices, and 
$P =(\mbox{det} A_{1}-1) (\mbox{det} A_{2} -1)$.
In the case of two-mode thermal squeezed states with 
(diagonal) $4 \times 4$ correlation matrices $A_{i} = D_{i}$
with $i =1,2$, the fidelity was expressed  as \cite[eq. (34)]{paraoanu}
\begin{equation}
F(\rho_{1},\rho_{2}) = \sqrt{ \mbox{det} ({2 \over (A_{1} A_{2} +I)
 - \sqrt{(A_{1}^2 -I) (A_{2}^2-I)}})}
\end{equation}
which is the product of the fidelities of the corresponding
one-mode thermal states.

It would, then, be of interest, proceeding along such
 lines, to extend the
work of Twamley \cite{twamley} to obtain the Bures (minimal monotone)
metric on the {\it two}-party Gaussian states. The volume element of the
metric could, then, serve as an {\it a priori} measure on these states.
We shall seek to do this below but first
  it seems quite relevant  to 
assess the relative separability/inseparability of these Gaussian
states by proceeding in a purely 
classical/nonquantum fashion. (However, in the present absence of explicit
formulas for the Bures metric on the two-party Gaussian states, we must
rely upon numerical methods.)
\section{CLASSICAL ANALYSIS}

 In sharp
 contrast to the quantum state-of-affairs \cite{petzsudar}, classically there
is a {\it unique} monotone metric, given by the Fisher information
 (cf. \cite{frieden}).
The Jeffreys' prior \cite{kass} (of Bayesian theory \cite{bernardo}) is, then,
taken to be proportional to the volume element of the Fisher information
metric. For the $n$-dimensional Gaussian (multivariate normal) probability
distributions (with fixed vectors of means), Jeffreys' prior is inversely
proportional to the ${n+1 \over 2}$-power of the determinant of the
$n \times n$ covariance matrix \cite{press}. (This determinant is invariant
under symplectic transformations. ``The symplectic group acts unitarily and
irreducibly on the two-mode Hilbert space'' \cite{simon}.)

To examine the separability properties of the two-party Gaussian states,
{\it vis-\'a-vis} this distinguished classical measure, we randomly generated
$4 \times 4$ real symmetric matrices 
($M$), using uniform random variates drawn from
 intervals of the form  $[0,k]$ for the four diagonal entries of $M$
 and  intervals
of the type
$[-l,l]$ for the six distinct off-diagonal entries
of $M$. (In an extensive survey article, Holmes \cite{holmes} has 
reviewed methods for generating random {\it correlation} matrices, that is
covariance matrices with all their diagonal entries equal to unity.
Let us note, however, that in the physics literature, the terminology 
``correlation matrix'' usually includes what statisticians would more
broadly call a variance/covariance matrix.)
 We, then, tested each
such matrix for positive definiteness. If it proved to be positive definite
(that is, its four eigenvalues were positive), we accepted it as a covariance
matrix of a (zero-mean four-vector) Gaussian
 probability distribution. We, then, further
tested $M$  to see that (in the notation of \cite[eq. (8)]{duan}),
the conditions $n, m  \geq 1$ on the entries of the ({\it locally} unitarily) 
transformed
Gaussian state were satisfied. If they were, we then examined
if the quantity expressed
in  the form $|a_{0}^2-{1 \over a_{0}^2}|$ was exceeded by the total
variance of the pair of Einstein-Podolsky-Rosen-type
operators stipulated by Duan {\it et al}, as required by the
uncertainty relation
(cf. \cite{kruger}). (The determination of 
the parameter $a_{0}$ requires the solution of a 
pair of nonlinear simultaneous equations. Simon \cite{simon} suggests that, 
in this respect, his implementation of the Peres-Horodecki separability
criterion is simpler than that of Duan {\it et al}.) 
Additionally, we checked (using inequality (16) of \cite{duan}) 
if the total variance
exceeded the ``much stronger bound'' $a_{0}^2 + {1 \over a_{0}^2}$,
 necessary and
sufficient for the separability of the associated Gaussian two-party state.

As an example (Table~\ref{table1}),
 for the choices $k=15$, $l=15$, we generated ten million
 $4 \times 4$ real symmetric
matrices ($M$).
 Of these, 39,588 fulfilled the positive definiteness requirement and
the inequality conditions (pertaining to the 
principal minors of $M$)
 on $n$ and $m$, and passed the test based on
the uncertainty relation. Of those, 39,003 exceeded the 
(more demanding) separability
criterion. Assigning the $-{5 \over 2}$-power of the 
determinant of the covariance matrix
as the (Fisher information) weight to each of the 39,588  matrices,
and normalizing by the sum of  all these  weights,
we found the relative probability of separability to be .991365.
Of the 39,003 separable states, 20,534 
still were associated with  positive definite matrices
after subtraction of the identity matrix from
their covariance matrices. They, therefore, possess positive
Glauber-Sudarshan 
$P$-representations \cite[sec. 3]{scutarunew}. (In the notation of
Simon \cite{simon}, one subtracts {\it one-half}
 times the identity matrix from the covariance matrix
for the test of classicality, but he employs a
different set of commutation relations from those of Duan {\it et al}.)
 The probability assigned to these
20,534 classical states was .000003027. Therefore, the probability
allocated to the separable states was mostly concentrated on the 
18,469 states that were nonclassical.
It is of interest, in this context, to note that the argument of Duan
{\it et al} relies upon the transformation using local linear unitary
Bogoliubov operations (LLUBO's) of the covariance matrix associated with
a separable state to a covariance matrix associated with a classical state
(that is, one possessing a positive $P$-representation).
Our results, then, clearly indicate that the property of possessing a 
positive
$P$-representation is not necessarily preserved under the inversion of these 
LLUBO's. (Gardiner \cite[p. 328]{gardiner}
 leaves as an exercise to show ``that the
squeezed state {\it never} has a Glauber-Sudarshan $P$-function for any
non-zero $\zeta$. (It does have of course, a positive $P$-function.)'')

We have repeated this  form of analysis for other choices of
$k$ and $l$, as well. These results are reported in Table~\ref{table1}.
The thrust of the results is much the same, although the probability of
separability is significantly  reduced in the $k=30$, $l =20$ analysis.
(We are not cognizant of any particular rationale for this 
observed reduction.)
Rather remarkably, the small probabilities of finding a state with a positive
Glauber-Sudarshan
$P$-representation all fall within a quite narrow range of probabilities,
for the different selections of $k$ and $l$.

\section{QUANTUM-THEORETIC ANALYSIS}

We would now like to repeat the form of analysis above, but using instead
of the volume element of the Fisher information matrix as a measure,
the volume element of the Bures 
(minimal monotone) metric \cite{braunstein,hubner}.
 To proceed, we are able, using the
detailed prescriptions in \cite{simonsudar}, in particular, eqs. (3.1), (3.9)
and (3.10) there, to construct (for a choice of $k$ and $l$)
the Gaussian form of 
the Schr\"odinger kernel for each (uniformly randomly generated covariance
 matrix of a Gaussian two-party quantum state)
$M$. (Note that $G$, in the notation of \cite{simonsudar}, is the same as
$M^{-1}$.)
Then, we approximate this nonnegative-definite
 kernel by $m^2 \times m^2$ matrices
($\Gamma$), for various
choices of odd $m$. For the values to substitute into the kernel, we used
the intersection points of a regular
square $m \times m$ lattice, with
unit spacing, centered at the 
origin.
We computed the $m^2$ eigenvalues 
($\lambda$'s) of  $\Gamma$, and, if they were all positive, normalized them to
exactly sum to unity. (If not, we 
immediately proceeded onto the next randomly
generated  matrix.) As the volume
element ($V_{Bures}$) of the Bures metric, we employed \cite[eq. (2)]{slater2}
\begin{equation} \label{approxvol}
V_{Bures} \propto (\mbox{det}{\Gamma})^{-{1 \over 2}}
 \prod_{1 \leq i <  j \leq m^2} {1 / (\lambda_{i} + \lambda_{j})} .
\end{equation}
(Let us observe that Kr\"uger \cite{kruger} has presented a diagonalization
of {\it one}-mode Gaussian states, based on Mehler's formula, giving a 
bilinear generating function for the Hermite polynomials. It should be
possible to extend this approach to the two-party Gaussian states, making
use of multidimensional Hermite polynomials \cite{dodonov}. In particular,
Louck  \cite{louck} has given a multivariate extension of Mehler's
formula (cf. \cite{slepian,holmquist,viskov}).)
The formula (\ref{approxvol}) is derived from the general formula for the
(infinitesimal) Bures distance \cite{hubner}
\begin{equation}
d_{Bures}^{2}(\rho,\rho+ \mbox{d} \rho) = \sum_{i,j=1}^{n} {1 \over 2} 
{{| <i| \mbox{d} \rho | j >}^{2} \over \lambda_{i} + \lambda_{j}},
\end{equation}
where $|i>$ denotes the eigenvectors of the $n \times n$ density matrix
$\rho$ and $<j|$ the corresponding complex conjugate (dual) vectors.

For $k=15$, $l=15$ and $m = 3, 5, 7, 9, 11$, we ran extensive analyses.
All these yielded probabilities of separability and classicality
essentially unity and zero, respectively.
However, since in the computation of the Bures volume (\ref{approxvol}),
long strings of products are computed, 
having very large and small outcomes, we did not have as much confidence
in the results, from a numerical point of view, 
as we would hope. Therefore, we sought, additionally, to pursue
an approach which it was believed should 
be rather {\it robust} \cite{huber} in this regard. We chose the parameters
$k=15$, $l=15$ and $m=5$. But now rather than using a square $25 \times 25$
grid, we used {\it five} rectangular (but not, in general,
 square) $25 \times 25$ grids.
The five points (set to be the same along both axes), were now chosen 
{\it randomly}
from the interval [-2,2]. Then,  with each of these five 
random rectangular grids, we associated
a $25  \times 25$ matrix, by substituting the values at the intersection
points of the grids 
into the Gaussian kernel, as described in 
\cite{simonsudar}. The eigenvalues ($\lambda$'s) and the Bures volumes 
(\ref{approxvol}) were
computed for each of the five cases. (If any of the eigenvalues
were nonpositive, we 
immediately discarded the present randomly generated covariance
matrix from consideration and proceeded onto the examination of 
the next one.)
 The five volume elements were ordered by magnitude, and we
chose for our subsequent computations of probabilities 
of separability and classicality, two robust estimators of the true
volume.
 One was the {\it median} of the five figures, and the
other was the {\it trimmed mean},
 obtained by averaging the second, third and fourth
largest values (thus, of course, fully ignoring the smallest and largest 
estimates of the volume element).

Following this scheme, we generated 18,600,000 random real symmetric
matrices. Of these, 74,431 were positive definite and did 
not violate the uncertainty relation.
From them, we discarded 4,943, since negative eigenvalues appeared in at least
one of the five $25 \times 25$ matrices generated from the random grids.
Of the remaining 69,488, there were 68,952 that were separable and 38,620
that were classical in nature.
 Applying the two (median and trimmed mean)  robust estimators
of the Bures volumes (\ref{approxvol}) as weights to these covariance matrices,
we obtained 
for both estimators 
(to many places of precision), probabilities of separability
equal to unity and  probabilities of classicality that were zero.
So, our robust calculations simply served to reconfirm our earlier
(nonrobust) ones.
In parallel with these computations, we recomputed the two probabilities, 
using the Fisher information metric. The probability of separability was now
.86397 and that of classicality, .00968. We suspect that the rather noticeable
changes from Table I are a consequence of the necessity to discard roughly
seven percent of the covariance matrices. But we would think that if this
discarding also affects the results based on the Bures metric, it should
tend to 
similarly both decrease the probability of separability and increase that of
classicality, effects, which  if present at all, are imperceptibly small in
nature.

In a study \cite{slater1}
 of {\it finite}-dimensional quantum systems, 
analogous to the investigation here (of infinite-dimensional systems),
Slater found that the semiclassical probability of separability 
(based, in part, on the Fisher information metric) of the
$2 \times 2$ systems was approximately thirty-six 
percent, while the quantum-theoretic (Bures/minimal monotone)
probability was roughly ten percent. The use of other monotone metrics,
such as the Kubo-Mori and maximal ones \cite{petzsudar},
 led to further reductions in this
probability. (Of the monotone metrics, the Bures/minimal one 
 has been found to be the least {\it noninformative} and the maximal 
monotone metric, 
the most \cite{slatercomp}.) Proceeding  along such lines, we
repeated our form of analysis for $k=15$, $l=15$, $m=7$, based on the single
uniformly-spaced square grid for both these alternative forms of metrics.
This involved replacing the term $ ( \lambda_{i} + \lambda_{j} )$, which is
of course proportional to the {\it arithmetic} mean of the two
eigenvalues,  in
(\ref{approxvol}) by $( \lambda_{i} - \lambda_{j} ) / ( \log{\lambda_{i}}
- \log{\lambda_{j}} ) $  which is proportional to
the {\it logarithmic} mean, for the Kubo-Mori case, and by
$ \lambda_{i} \lambda_{j}  /( \lambda_{i} + \lambda_{j} ) $,
which is proportional to the {\it harmonic} mean, for the maximal
monotone scenario. However, our initial analyses produced numerical errors
(divisions by zero), and we have not yet pursued the matter further.

\section{CONCLUDING REMARKS}

\.Zyczkowski \cite{zycz2} has argued that if one controls for
 the participation ratio, $R(\rho) = 1/\mbox{Tr}(\rho^{2})$, of density 
matrices $\rho$, then, the probability of separability should
be essentially invariant for a broad class of natural measures on the
space of $\rho$'s, that is, ``the link between the purity of the mixed states
and the probability of entanglement is not sensitive to 
the measure chosen''. (The participation ratio is unity for a pure state, and
greater than unity for a mixed/impure state.)
The results reported here seem even stronger in this regard,  
since both the classical (Fisher information) and
quantum-theoretic 
(Bures) measures used  give quite similar results {\it in toto},
without the need to control for the degree of purity.

It would be of interest to examine the cases of the $n$-party
Gaussian states ($n > 2$), as well, in particular to ascertain if the
{\it a priori} probability of separability decreases from its apparent
high value for $n=2$, 
but neither sufficient nor necessary conditions
for separability appear to have been developed for them. 
Also, it
would be desirable to conduct a classical (Fisher information
metric) analysis
of the {\it one}-mode thermal squeezed (Gaussian) 
states. This result could then be compared
with the
quantum-theoretic one
 of Twamley \cite{twamley}, based on the Bures (minimal monotone)
metric.

As a possible caveat here, it should be pointed out that in \cite{slater2},
for the squeezed thermal states, the one-dimensional marginal $g(\beta)$
(\ref{impropbures}) of the Bures volume element 
was found to be improper/unnormalizable over the full range of
possible values of $\beta \in [0,\infty]$, as well as $f(r)$ over
$r \in [0,\infty]$. This would strongly suggest that
the volume element of the Bures metric (for which no explicit formula is
presently available) for the Gaussian two-party states is also improper.
(A similar conclusion would appear to hold in terms of  Jeffreys'
prior.)
Therefore, in some strict sense, it may be inappropriate to speak in terms
of the probability of separability or classicality 
for the Gaussian two-party states {\it in toto}, but rather one would have to
restrict oneself to some subset of these states, for which the corresponding
volume element is, in fact, normalizable.

Although, of course, for the  Gaussian one-party states, the concept of
separability is not applicable, one may still inquire concerning the
associated 
 probability of classicality. We have investigated this question, using
the  Jeffreys' prior, and were able to determine
 formally that in the limit as one
considers all Gaussian one-party states, the probability of classicality 
 converges to  zero. This, of course, conforms strongly to our results for the 
Gaussian two-party states.

We would also like to point out the interesting question of whether or not
the necessary separability (entropic) criterion of Cerf and Adami
\cite{cerf} carries over from discrete to continuous systems. Our
preliminary analyses, in the context of this study, using the general 
formula for the entropy of a quantum Gaussian state \cite{holevo1,holevo2}, 
appear to indicate that
it does {\it not}. This may not be altogether surprising, as ``one may even 
claim that discrete entropies and continuous entropies are fundamentally 
different in their nature and, to some extent, in their meanings'' 
\cite[p. 33]{jumarie} (cf. \cite{rajagopal}).

The statistical/Bayesian interpretation of the main results 
here would appear to be
that it should
 take a very considerable amount of evidence (that is, a large number of
repeated measurements) to convince one that an 
initially unknown Gaussian two-party state is, in fact, inseparable or
classical in nature, the {\it a priori} assumption being (in the absence of
any specific information available) that the state is, in all probability, both
separable (classically correlated) and nonclassical.

\acknowledgments

I would like to express appreciation to the Institute for Theoretical
Physics for computational support in this research.

\begin{table}
\caption{Probabilities of separability and classicality for Gaussian
two-party states based on the volume element of the 
Fisher information metric, that is, Jeffreys' 
prior, for various choices
of $k$ and $l$. These parameters determinine the ranges,
$[0,k]$ and $[-l,l]$,  of the
diagonal and off-diagonal entries, respectively,
 of uniformly randomly generated real
symmetric
matrices. If 
 these matrices are positive definite and do not violate the
uncertainty relation, their 
aggregate separability and classicality properties are
studied and tabulated here.}

\label{table1}
\begin{tabular}[p]{l  | | r r  | |  r r r  | |  r r}
 random matrices & $k$ & $l$ & insep. and  sep.
 & separable  &  classical  & $prob_{sep}$
 & $prob_{classical}$ \\
\hline
500,000 & 10 & 5 & 58,835 & 57,663 & 27,646 & .993330 & .000001470 \\
1,900,000 & 500  & 250 & 119,447 & 119,447 & 116,781 & 1.000000 & .000004259 \\
5,200,000 & 20 & 10 & 650,718 & 648,475 & 451,820 & .998716 & .000003094 \\
8,100,000 & 30 & 20 & 316,319 & 315,622 & 235,459 & .938819 & .000001701 \\
10,000,000 & 15 & 15 & 39,558 &  39,003 & 20,534 & .991365 & .000003027 \\
\hline
\end{tabular}
\end{table}

\end{document}